\documentclass[aps,prl,twocolumn,letterpaper,superscriptaddress,10pt]{revtex4-2}
\usepackage{amssymb,amsthm,amsmath,amsfonts}
\usepackage{graphicx,ulem,mathptmx}
\usepackage[pdftex,dvipsnames,usenames]{xcolor}
\usepackage[colorlinks=true,urlcolor=blue,citecolor=blue,linkcolor=blue]{hyperref}
\usepackage{comment,bm}
\usepackage[shortlabels]{enumitem}

\begin{document}


\title{Direct Measurement of Higher-Order Nonlinear Polarization Squeezing}

\author{Nidhin Prasannan}
	\affiliation{Integrated Quantum Optics Group, Institute for Photonic Quantum Systems (PhoQS), Paderborn University, Warburger Stra\ss{}e 100, 33098 Paderborn, Germany}
\author{Jan Sperling}
	\affiliation{Theoretical Quantum Science, Institute for Photonic Quantum Systems (PhoQS), Paderborn University, Warburger Stra\ss{}e 100, 33098 Paderborn, Germany}
\author{Benjamin Brecht}
	\affiliation{Integrated Quantum Optics Group, Institute for Photonic Quantum Systems (PhoQS), Paderborn University, Warburger Stra\ss{}e 100, 33098 Paderborn, Germany}
\author{Christine Silberhorn}
	\affiliation{Integrated Quantum Optics Group, Institute for Photonic Quantum Systems (PhoQS), Paderborn University, Warburger Stra\ss{}e 100, 33098 Paderborn, Germany}

\begin{abstract}
    We report on nonlinear squeezing effects of polarization states of light by harnessing the intrinsic correlations from a polarization-entangled light source and click-counting measurements.
    Nonlinear Stokes operators are obtained from harnessing the click-counting theory in combination with angular-momentum-type algebras.
    To quantify quantum effects, theoretical bounds are derived for second- and higher-order moments of nonlinear Stokes operators.
    The experimental validation of our concept is rendered possible by developing an efficient source, using a spectrally decorrelated type-II phase-matched waveguide inside a Sagnac interferometer.
    Correlated click statistics and moments are directly obtained from an eight-time-bin quasi-photon-number-resolving detection system.
    Macroscopic Bell states that are readily available with our source show the distinct nature of nonlinear polarization squeezing in up to eighth-order correlations, matching our theoretical predictions.
    Furthermore, our data certify nonclassical correlations with high statistical significance, without the need to correct for experimental imperfections and limitations.
    Also, our nonlinear squeezing can identify nonclassicality of noisy quantum states which is undetectable with the known linear polarization-squeezing criterion.
\end{abstract}

\date{\today}
\maketitle

\paragraph*{Introduction.---}

    Squeezing plays a vital role as a fundamental quantum effect and fosters today's development of quantum technologies.
    Squeezed states of light are studied on various platforms \cite{SHYMV85, WKHW86,HFTKTS05,FSKPW02,GSYL87}, harnessing their nonclassical properties to advance the performance of imaging, sensing, and information processing applications \cite{PCK92,Ligo16,Aetal13,H00,VMDS16,TJDKHBB13}.
    Among the different ways squeezing can manifest itself, spin and polarization squeezing of light are of particular interest \cite{WCBPKM10,HSDM12,LJKM16,BZYCQHWW22}.

    In classical (statistical) optics, polarization properties are explained via Stokes parameters, visualized on the Poincar\'e sphere.
    For quantum light, the operator counterparts to Stokes parameters were established \cite{GHBKGLS21}.
    Similar to squeezing in phase space, accessible via quadrature (i.e., field) operators, the idea here is to assess squeezing in terms of Stokes-operator fluctuations \cite{AC03}.
    Different theoretical techniques to quantify spin squeezing were proposed \cite{K97,KLLRS02,ASST02}, and a number of experiments have been performed \cite{HJLA05,BSBL02}.

    Experiments often employ second-order moments for probing the quantum-classical boundary, not appreciating the information content provided by higher-order correlations;
    see, e.g., Ref. \cite{ALCS10} for one exception.
    For quadrature squeezing, higher-order effects have been considered \cite{HM85,H87}.
    Indeed, for characterizing the quantum polarization of light, it was shown that higher-order polarization properties are essential \cite{SBKSL12}.
    In addition, and beyond the linear regime, it is exceedingly interesting to study nonlinear quantum phenomena that are inaccessible via simple linear functional dependencies alone \cite{K97}.
    However, nonlinear quantum effects are hard to detect at best and often simply unattainable.

    The detection of complexly structured quantum light is challenging.
    In particular, phase-sensitive measurements typically require a well-defined external reference phase, such as provided by the local oscillator in balanced homodyne detection \cite{SBRF93}.
    But, for polarization measurements, interference properties of the two polarization components suffice \cite{AC03,GHBKGLS21}.
    Still, such measurements commonly require photon-number-resolution capabilities, which are generally not available.
    Consequently, pseudo-number-resolving detection has been established to mitigate such limitations \cite{PTKJ96,ASSBW03,FJPF03,BDFL08}.
    Using the therefore developed click-counting framework \cite{SVA12}, moment-based criteria allow for detecting nonclassical features, even with incomplete photon-number information \cite{SVA13,SBVHBAS15}.
    Nowadays, large systems of up to 128 time-bin-multiplexed detectors are available, allowing us to explore macroscopic quantum correlations \cite{JEBSC21}.
    While first theoretical attempts were made to combine click counting with nonclassical polarization states \cite{SVA16}, a full theoretical description and an experimental demonstration were not realized to date.

    Polarization-entangled sources based on parametric down-conversion have been successfully investigated \cite{KMWZSS95,WCWBHSVANPSP16}.
    Yet, these experiments often solely exploit single-photon components to produce entangled Bell and GHZ states \cite{MPDEQBBPS19} in a low-pump-power approximation.
    Intrinsically, however, the down-conversion process also leads to higher-order contributions.
    And a full expansion yields a macroscopic Bell state \cite{IACL12}, i.e., a continuous-variable Einstein--Podolski--Rosen state, whose quantum noise properties have been characterized in terms of second-order, linear Stokes-operator fluctuations \cite{ICRL11};
    however, nonlinear functional dependencies remain unexplored.

    We develop a nonlinear Stokes-operators formalism in terms of click-counting theory.
    Based on this approach, second- and higher-order moment-based inequalities are derived that, when violated, certify nonclassical polarization states. 
    We combine an efficient source of entangled light with a click-counting detection unit for the experimental verification of nonclassicality.
    A parametric down-conversion source in a Sagnac loop \cite{MPWQDBS18} allows us to produce macroscopic Bell states with higher-order photon-number contributions and different polarization features, including complete and partial polarization nonclassicality.
    From the recorded click pattern, we then directly reconstruct up to eighth-order moments for nonlinear Stokes operators, characterizing nonlinear quantum polarization effects with high statistical significance and without corrections for measurement imperfections.

\paragraph*{Theory.---}

	We employ a detection scheme in which incident light is split into $N=8$ bins of equal intensity, each measured with a single-photon detector \cite{PTKJ96,ASSBW03,FJPF03,BDFL08}.
	It was shown that obtaining $k$ nonvacuum signals, i.e., clicks, is described by a positive operator-valued measure that exhibits a binomial form in a normally ordered expansion, $\hat\Pi_k={:}\binom{N}{k}\hat\pi^k(\hat 1-\hat\pi)^{N-k}{:}$ \cite{SVA12}, with $\hat \pi=\hat 1-{:}\exp(-\eta \hat n/N){:}$.
	Therein, $\eta$, $\hat n$, and ${:}\cdots{:}$ are the detection efficiency, the photon-number operator, and the normal order symbol, respectively.
	Importantly, this click-counting description is different from the common Poisson form for a full photon-number resolution, $\hat\Pi_k={:}e^{-\eta\hat n}(\eta\hat n)^k/k!{:}+\mathcal{O}(1/N)$, only slowly $1/N$ converging toward the photoelectric Poisson model \cite{SVA12}.

	For the two quantized polarization components, the linear representation is given in terms of Stokes operators.
	(See, e.g., Refs. \cite{CMMSZ06,BSSKGMM10} for introductions.)
	That is, a wave-plate-based transformation of, say, horizontal and vertical photon numbers results in $\hat a^\dag\hat a\mapsto(\hat S_0+\boldsymbol{e}\cdot \boldsymbol{\hat S})/2$ and $\hat b^\dag\hat b\mapsto(\hat S_0-\boldsymbol{e}\cdot \boldsymbol{\hat S})/2$, respectively.
	Therein, the Stokes-operator vector is given by $\boldsymbol{\hat S}=(\hat a^\dag\hat b+\hat b^\dag\hat a,-i\hat a^\dag\hat b+i\hat b^\dag\hat a,\hat a^\dag\hat a-\hat b^\dag\hat b)$, the total photon number reads $\hat S_0=\hat a^\dag\hat a+\hat b^\dag\hat b$, and the vector $\boldsymbol{e}=(\sin\vartheta\cos\varphi,\sin\vartheta\sin\varphi,\cos\vartheta)$ corresponds to the measurement projection direction on the Poincar\'e sphere.
	For the applied combination of quarter-wave (QWP) and half-wave (HWP) plates, the relative phase is $\varphi=\arg(\rho^\ast\tau)\in[0,2\pi[$, the transmission coefficient is $|\tau|=\cos\vartheta/2$ ($0\leq \vartheta\leq \pi$), and the reflection coefficient is $|\rho|=\sin\vartheta/2$.

	In our scenario, however, the mean click number for the two polarizations is $\langle N \hat\pi_{A}\rangle=N(1-\langle{:}\exp[-\eta (\hat S_0+\boldsymbol{e}\cdot \boldsymbol{\hat S})/(2N)]{:}\rangle)$ and $\langle N \hat\pi_{B}\rangle=N(1-\langle {:}\exp[-\eta (\hat S_0-\boldsymbol{e}\cdot \boldsymbol{\hat S})/(2N)]{:}\rangle)$ \cite{SVA16,SVA13}.
	To characterize the polarization state, one typically utilizes the difference photon number, $\hat a^\dag\hat a-\hat b^\dag\hat b\mapsto \boldsymbol{e}\cdot \boldsymbol{\hat S}$, resembling a detection on Poincar\'e sphere along $\boldsymbol{e}$.
	Analogously, we here consider the mean difference of clicks, given by the expectation value of
	\begin{equation}
		\label{eq:StokesNL}
		\hat S_\mathrm{NL}=N\hat\pi_A{-}N\hat \pi_B
		=2N{:}\exp\left(-\frac{\eta}{2N}\hat S_0\right)
		\sinh\left(\frac{\eta}{2N}\boldsymbol{e}\cdot \boldsymbol{\hat S}\right){:}.
	\end{equation}
	Importantly, this operator is a nonlinear function of Stokes operators which includes a hyperbolic sine of the sought-after projection $\boldsymbol{e}\cdot \boldsymbol{\hat S}$, and which also includes an exponential scaling with the total photon number ($\hat S_0$), accounting for detector saturation.
	This nonlinear nature is an intrinsic feature when combining modern click-counting theory with angular-momentum algebras, superseding the aforementioned linear counterpart.
	Note that the limit $N\to\infty$ yields $\hat S_\mathrm{NL}=\boldsymbol{e}\cdot \boldsymbol{\hat S}+\mathcal{O}(1/N)$, and Eq. \eqref{eq:StokesNL} relates to single-mode balanced homodyne detection with click counting \cite{SVA15}.

	Similarly to the difference, the total (i.e., summed) click-number operator, mirroring the total photon number, reads
	\begin{equation}
	    \label{eq:Stokes0NL}
	\begin{aligned}
		&\hat S_\mathrm{0,NL}=N\hat\pi_A+N\hat \pi_B
		\\
		=&2N\left[\hat 1-{:}\exp\left(-\frac{\eta}{2N}\hat S_0\right)
		\cosh\left(\frac{\eta}{2N}\boldsymbol{e}\cdot \boldsymbol{\hat S}\right){:}\right]
	\end{aligned}
	\end{equation}
    and additionally depends on $\boldsymbol{e}\cdot \boldsymbol{\hat S}$, contrasting the linear total photon number $\hat S_0$ without similar contributions;
    see Ref. \cite{LSV15} for related considerations for homodyne detection.

	For determining polarization nonclassicality, we can now use the known method of normally ordered matrices of moments \cite{AT92,SRV05}, applied here to nonlinear Stokes operators.
	The resulting matrix $M=(\langle{:}S_\mathrm{NL}^{k+l}{:}\rangle)_{k,l=0,\ldots,N/2}$ includes moments up to the $N$th order and is positive semidefinite for classical light, $M\geq 0$ \cite{SVA13}.
	If, however, $M\ngeq 0$ applies, nonclassicality is certified.
	As one example, we can consider the determinant of the principal leading two-dimensional submatrix of $M$.
	Then, the second-order, classical constraint reads
	\begin{equation}
	    \label{eq:2ndOrderFluctuations}
		0\stackrel{\text{cl.}}{\leq}\det\begin{pmatrix}
			\langle{:} \hat S_\mathrm{NL}^0{:}\rangle & \langle{:} \hat S_\mathrm{NL}^1{:}\rangle
			\\
			\langle {:}\hat S_\mathrm{NL}^1{:}\rangle & \langle{:} \hat S_\mathrm{NL}^2{:}\rangle
		\end{pmatrix}
		=\langle {:}(\Delta\hat S_\mathrm{NL})^2{:}\rangle,
	\end{equation}
	with $\langle{:} \hat S_\mathrm{NL}^0{:}\rangle=\langle \hat 1\rangle=1$.
	This inequality lower bounds the normally ordered and nonlinear variance of $\hat S_\mathrm{NL}$ for classical polarization states by zero.
	It is also worth mentioning that the elements of $M$ can be directly expressed through moments of $\hat\pi_A$ and $\hat\pi_B$ through Eq. \eqref{eq:StokesNL} which we obtain from $\langle{:}\hat \pi_A^{j}\hat \pi_B^{j'}{:}\rangle=\sum_{k=j}^N\sum_{l=j'}^N c_{k,l} \binom{k}{j}\binom{l}{j'}\binom{N}{j}^{-1}\binom{N}{j'}^{-1}$ \cite{SBVHBAS15}, solely utilizing the actually measured joint click-counting statistics $c_{k,l}$.

	In the Supplemental Material (SM) \cite{SuppMat}, we show that our second-order nonlinear squeezing criterion \eqref{eq:2ndOrderFluctuations} is more noise resilient than its linear conterparts.
	This result is obtained by considering polarization-entangled single photons with thermal background as examples which exhibit no linear, but nonlinear squeezing.
	For instance, for $N=8$, we can accept more than $50\%$ more noise with our nonlinear Stokes formalism.

\paragraph*{Experiment.---}

\begin{figure}[t]
    \includegraphics[width=\columnwidth]{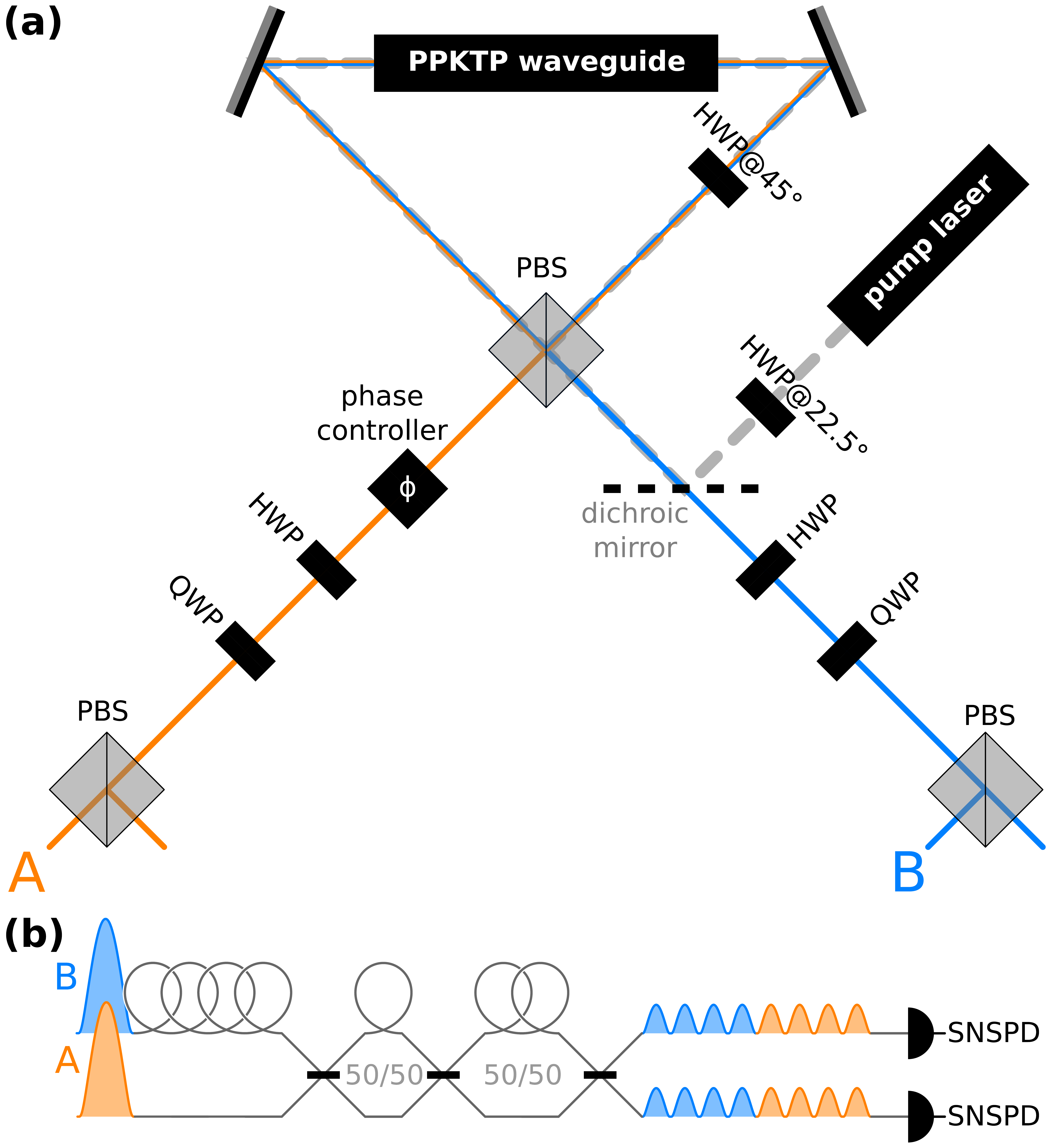}
    \caption{
        Setup outline.
        A $775~\mathrm{nm}$ laser pumps the waveguide source with a spectrally filtered pump width of $1~\mathrm{ps}$.
        Polarization-entangled photon pairs are generated in the forward and backward propagation in the Sagnac interferometer and separated with a polarizing beam splitter (PBS).
        A phase controller allows us to switch between symmetric and antisymmetric Bell states.
        Combinations of a HWP and QWP with a PBS yield arbitrary Stokes projections $\boldsymbol{e}\cdot \boldsymbol{\hat S}$ on the Poincar\'e sphere.
        Our time-multiplexed detector (TMD) is built from low-loss fibers and $50/50$ beam splitters. 
        The resulting time-bin-resolved photons in arms $A$ and $B$ are detected by two superconducting nanowire single-photon detectors (SNSPDs) with an efficiency exceeding $80\%$.
    }\label{fig:setup}
\end{figure}

\begin{figure*}
    \includegraphics[width=\textwidth]{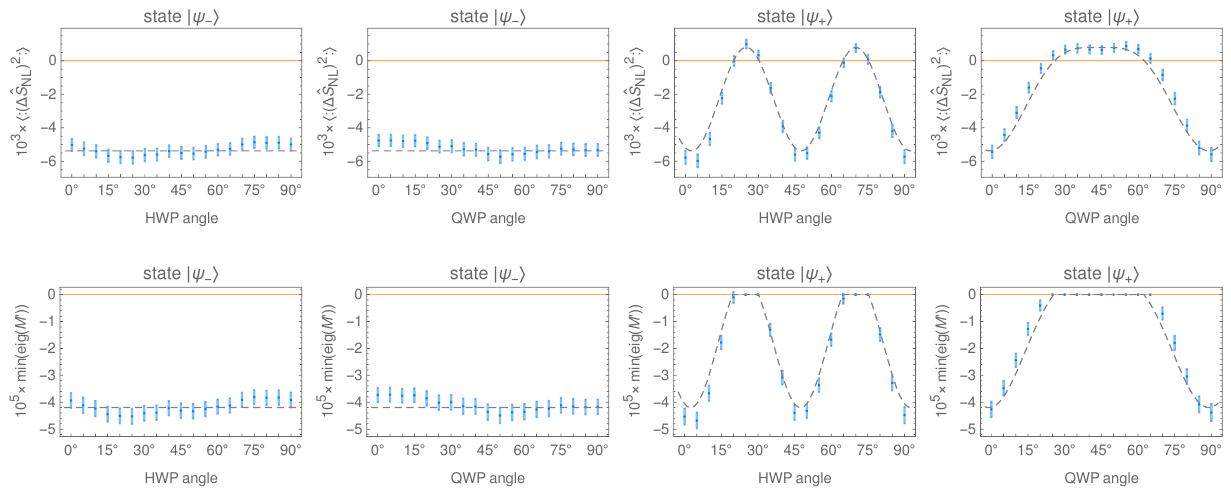}
    \caption{
        Second-order and higher-order nonlinear polarization nonclassicality are depicted as negativities in the top and bottom row of plots, respectively.
        The first column of plots shows the results for the symmetric macroscopic Bell state in Eq. \eqref{eq:MacroBell} for $\phi=0$ as a function of the QWP angle and a fixed HWP angle at zero degree, and vice versa for the second column.
        The third and fourth columns include analogous findings for the antisymmetric macroscopic Bell state, $\phi=\pi$.
        A $\pm15\sigma$ error margin is provided as a vertical bar.
        The theoretical model is shown as dashed lines in all plots for the single set of fit parameters $\lambda=0.36$ and $\eta=0.135$.
    }\label{fig:results}
\end{figure*}

    See Fig. \ref{fig:setup}(a) for the setup and Fig. \ref{fig:setup}(b) for the click detection.
    See also SM \cite{SuppMat} for additional details.

    A periodically poled potassium titanyl phosphate (PPKTP) waveguide is operated inside a Sagnac interferometer \cite{MPWQDBS18}.
    Bidirectional pumping efficiently generates type-II parametric down-conversion in clockwise and counterclockwise propagation direction with non-negligible higher-order contributions. 
    The pump light is delivered from a pulsed laser with $100~\mathrm{fs}$ pulse duration and $774~\mathrm{nm}$ central wavelength.
    A folded $4f$ spectrometer is used to select a narrow part of the pump spectrum, yielding pump pulses with $1~\mathrm{ps}$ duration.
    Our engineered waveguide source generates spectrally decorrelated photon pairs independently in both directions.
    A HWP at a fixed angle, $45^\circ$, inside the interferometer switches the polarization, ensuring that pump and down-converted (i.e., signal and idler) photons have identical polarization when they recombine.
    Interfering the resulting beams on the Sagnac polarizing beam splitter (PBS) generates entangled light in two output modes.
    The source performance is benchmarked with a visibility $>96\%$ and a fidelity $>95\%$ in the low-pump-power regime.
    The phase in arm $A$ is tuned with a $1550~\mathrm{nm}$ Soleil--Babinet compensator, labeled as phase controller in Fig. \ref{fig:setup}(a).
    HWP and QWP combinations are used to perform arbitrary Stokes measurements.
    The measurement PBSs project the state into the two polarization modes and direct beams $A$ and $B$ to our detection unit.

    For click counting, the produced light is sent through an eight-time-bin detection unit, Fig.\ref{fig:setup}(b).
    Our time-multiplexed detector (TMD) consists of a series of three $50/50$ beam splitters connected by delay fibers with different lengths, resulting in a splitting into different time bins \cite{ASSBW03,FJPF03}.
    Time bins are separated by $100~\mathrm{ns}$.
    Since two light beams are coming from the setup, both inputs of one TMD can be utilized in a delayed manner, allowing for a resource-efficient characterization.
    A total of 16 time bins for $A$ and $B$ are detected by two superconducting nanowire single-photon detectors (SNSPDs), which have dead time of $60~\mathrm{ns}$, much less than the bin separation.
    The repetition rate of the experiment is reduced to $1~\mathrm{MHz}$ to account for the time delays introduced by the TMD detection scheme.

\paragraph*{Results.---}

    Relatively strong pumping (here, $100~\mathrm{\mu W}$, or $0.1~\mathrm{nJ}$ per pulse) produces polarization-entangled states with higher-order photon-number contributions.
    Our waveguide-based approach benefits from a high degree of spatial confinement and, thus, produces bright quantum light, not limited as bulk crystal sources that suffer spatial distortions, e.g., Kerr lensing.
    The produced state after the phase controller reads
    \begin{equation}
        \label{eq:MacroBell}
        |\psi\rangle=(1-|\lambda|^2)
        \sum_{m,n=0}^\infty \lambda^m(e^{i\phi}\lambda)^n
        |m\rangle\otimes|n\rangle\otimes|n\rangle\otimes|m\rangle,
    \end{equation}
    with $|\lambda|<1$;
    the tensor products of number states are sorted according to output arm and polarization as follows: horizontal for $A$, vertical for $A$, horizontal for $B$, and vertical for $B$.
    We specifically use the controller settings $e^{i\phi}=\pm1$, defining macroscopic Bell states $|\psi_\pm\rangle$ \cite{ICRL11}.
    That name for the continuous-variable state \eqref{eq:MacroBell} originates from the two-photon subspace (i.e., $m+n=1$) in which we have an entangled two-qubit state---meaning that $|\psi\rangle\sim |H\rangle\otimes|V\rangle+e^{i\phi}|V\rangle\otimes|H\rangle$ for modes $A$ and $B$, with the horizontal and vertical single-photon states $|H\rangle=|1\rangle\otimes|0\rangle$ and $|V\rangle=|0\rangle\otimes|1\rangle$, respectively.
    (Note the symmetric and antisymmetric exchange symmetry between $A$ and $B$ for $e^{i\phi}=\pm 1$.)
    By controlling the squeezing strength via $\lambda$, one obtains entangled qubit and qudit states in the two-photon and few-photon regime, $|\lambda|\approx 0$, and macroscopic Bell states for high pump powers, i.e., higher $|\lambda|$.

    Using a two-mode polarization tomography, we collect data for $1\,300~\mathrm{s}$, resulting in ca. $10^8$ events with different coincidence counts $k$ and $l$ for $A$ and $B$, respectively, yielding the joint click-counting statistics $c_{k,l}$ for each waveplate setting and state $|\psi_\pm\rangle$.
    For simplicity, we vary either HWP or QWP angles, while keeping the other wave plate at $0^\circ$ and using the same angles for $A$ and $B$.
    From this data, we can directly infer the nonlinear Stokes-parameter fluctuations, Eq. \eqref{eq:2ndOrderFluctuations}, that are depicted in the top row of Fig. \ref{fig:results}.
    Via the highly significant negativities, nonlinear polarization squeezing is observed for the macroscopic Bell states.
    Please note that a 15-standard-deviation uncertainty is chosen in Fig.  \ref{fig:results} such that the error bars are actually visible, but not dominant in the plots.

    Our model (dashed lines in Fig. \ref{fig:results}) is based on our click-counting description in the theory part and the macroscopic Bell states in Eq. \eqref{eq:MacroBell} \cite{SuppMat}.
    Importantly, for all depicted scenarios, we use a single set of only two fit parameters.
    That is, the overall efficiency of our setup is estimated as $\eta=13.5\%$, and we have a relatively high $\lambda=0.360$---corresponding to $3.3~\mathrm{dB}$ quadrature squeezing.
    Small experiment-theory deviations originate from fluctuating pump powers over the experimental run, and the theoretical assumption that all polarization-altering components are perfectly aligned.

    Thus far, we demonstrated a nonlinear second-order nonclassical polarization, inequality \eqref{eq:2ndOrderFluctuations}.
    We also outlined that, with $N=8$ detection bins, a normally ordered matrix of up to eighth-order moments of $\hat S_\mathrm{NL}$ can be used.
    In addition, the total click number $\hat S_\mathrm{0,NL}$ is polarization-dependent, too, Eq. \eqref{eq:Stokes0NL}.
    Hence, we can use that information in terms of moments ${:}\hat S_\mathrm{0,NL}^m{:}$ (for $m=0,\ldots,N$) for our purposes as well, assigning a useful meaning to this operator.
    Because of relations \eqref{eq:StokesNL} and \eqref{eq:Stokes0NL}, we have $\hat\pi_A=(\hat S_\mathrm{0,NL}+\hat S_\mathrm{NL})/N$ and $\hat\pi_B=(\hat S_\mathrm{0,NL}-\hat S_\mathrm{NL})/N$.
    Thus, we can also use the following matrix that exploits all accessible moments of $\hat S_\mathrm{NL}$ and $\hat S_\mathrm{0,NL}$:
    \begin{equation}
        M'=(\langle{:}\hat\pi_A^{j_A+j'_A}\hat\pi_B^{j_B+j'_B}{:}\rangle)_{(j_A,j_B),(j'_A,j'_B)\in\{0,\ldots,N/2\}\times\{0,\ldots,N/2\}},
    \end{equation}
    where $N$ is even and rows and columns are defined through index pairs $(j_A , j_B )$ and $(j_A^\prime , j_B^\prime )$, respectively.
    (See Refs. \cite{SVA13,SBVHBAS15,LSV15} for further details.)
    If the minimal eigenvalue of $M'$ is below zero, $\min(\mathrm{eig}(M'))<0$, we infer up to eighth-order nonlinear polarization nonclassicality of $\hat S_\mathrm{NL}$ and $\hat S_\mathrm{0,NL}$, being the maximal information contents extractable from our measurement with $N=8$ detection bins \cite{SVA13,SVA16}.
    Beyond variance-based squeezing criteria, we can thereby explore nonclassical signatures in the skewness (third order), kurtosis (fourth order), etc. of nonlinear Stokes operators.
    Also, access to higher moments via more detection bins $N$ generally allows for an improved state reconstruction \cite{KSVS18}.

    The bottom row in Fig. \ref{fig:results} shows the results for our data for macroscopic Bell states $|\psi_\pm\rangle$.
    In all polarization settings, $|\psi_-\rangle$ shows a higher-order, nonlinear polarization well below the classical bound of zero.
    For $|\psi_+\rangle$, such nonclassical correlations depend on the wave plate settings.
    In comparison with the second-order approach, however, the higher-order moments never rise above the boundary at zero.
	In this context, it is important to understand that the relevant information lies in the relative values of negativities, and not their absolute magnitude.
	That is, an increased negativity in a particular waveplate configuration implies stronger quantum effects along this direction on the Poincar\'e sphere.
    Again, all higher-order nonclassical effects are certified with high statistical significance and agree with our theoretical prediction.

\paragraph*{Conclusion.---}

    We devised a theoretical approach for a nonlinear Stokes-operator framework that is directly accessible via state-of-the-art click-counting measurements.
    In addition, we derived criteria that render it possible to detect nonclassical polarization states of light in this manner.
    Our nonlinear Stokes-operator formalism has a significantly increased robustness against noise, rendering it possible to detect nonlinear polarization squeezing even when linear squeezing fails.
    Furthermore, we experimentally demonstrated nonlinear polarization squeezing and even higher-order nonclassical signatures.
    To this end, macroscopic Bell states were generated via a waveguide-based Sagnac source and measured using a single eight-bin click-counting detection unit for both polarizations.
    In contrast to measurements with true photon-number-resolving detectors, the total click number is also polarization dependent.
    We exploited this fact to certify nonlinear and nonclassical polarization by jointly using up to eighth-order moments for both nonlinear Stokes operators and total click number.
    All observed quantum signatures are in agreement with our theoretical model and are reported with high statistical significance and without requiring correction for imperfections, such as unavoidable losses, detector saturation, and incomplete photon-number resolution, all of which are included in the here-developed theory.

    Therefore, we formulated and implemented an easily accessible method to characterize the quantum nature of polarization states of light.
    This newfound potential can be harnessed in photonic quantum-enhanced applications, breaking classical bounds of what is achievable with linear polarization, ranging from qubits and qudits (i.e., few-photon states) all the way to macroscopic (i.e., continuous-variable) correlations.
    Our method may be relevant in future studies of macroscopic polarization entanglement \cite{MGM15} beyond nonclassicality, in quantum metrology and imaging applications \cite{DI92,YCCRB22,MAL19,CAGD19}, which have to be particularly noise resilient, and in studies of spin noise effects \cite{ORH05,AJA13,MJAA07} in the quantum domain.

\paragraph*{Acknowledgments.---}
    The authors thank Michael Stefszky for valuable comments.
    The Integrated Quantum Optics group acknowledges financial support through the European Commission through the H2020-FETFLAG-2018-03 project PhoG (Grant No. 820365) and the ERC project QuPoPCoRN (Grant No. 725366).
	J. S. acknowledges financial support from the Deutsche Forschungsgemeinschaft (DFG, German Research Foundation) through the Collaborative Research Center TRR 142 (Project No. 231447078, project C10).

\end{document}


\title{
	Direct Measurement of Higher-Order Nonlinear Polarization Squeezing:
	\\
	Supplemental Material
}

\author{Nidhin Prasannan}
	\affiliation{Integrated Quantum Optics Group, Institute for Photonic Quantum Systems (PhoQS), Paderborn University, Warburger Stra\ss{}e 100, 33098 Paderborn, Germany}

\author{Jan Sperling}
	\affiliation{Theoretical Quantum Science, Institute for Photonic Quantum Systems (PhoQS), Paderborn University, Warburger Stra\ss{}e 100, 33098 Paderborn, Germany}

\author{Benjamin Brecht}
	\affiliation{Integrated Quantum Optics Group, Institute for Photonic Quantum Systems (PhoQS), Paderborn University, Warburger Stra\ss{}e 100, 33098 Paderborn, Germany}

\author{Christine Silberhorn}
	\affiliation{Integrated Quantum Optics Group, Institute for Photonic Quantum Systems (PhoQS), Paderborn University, Warburger Stra\ss{}e 100, 33098 Paderborn, Germany}

\date{\today}

\begin{abstract}
	This supplemental documents collects additional information about theory and experiment for the results reported in the paper.
	The document is structured as follows:
	Further technical details on the experiment are provided in Sec. \ref{secSM:Experiment}.
	An auxiliary study that compares linear and nonlinear polarization-squeezing criteria on the basis of noise robustness can be found in Sec. \ref{sec:CompareLandNL}.
	Section \ref{secSM:Model} includes the exact calculations that are used to model our experiment and that are also the basis for fitting data and for making theoretical predictions.
\end{abstract}

\maketitle

\section{Experimental details}
\label{secSM:Experiment}

	In this section, we offer additional technical details on the experimental realization.
	This includes:
		a general setup characterization,
		measurements of the joint spectral intensity (JSI) of the nonlinear waveguide,
		filtering of the pump light,
		and a description time-multiplexing detector (TMD).

	A periodically poled $9\,\mathrm{mm}$-long potassium titanyl phosphate waveguide is used for the experiment.
	The source is engineered for an efficient, spectrally pure photon-pair production, at $1550\,\mathrm{nm}$ central wavelength. 
	The spectral behavior of the source is characterized using a time of flight spectrometer \cite{AEMS09}, where we recorded the JSI from photon arrival times through long dispersive fibers.
	Each pair photon is generated in a single temporal mode.
	From the JSI, we expect photon spectral purities of $98\%$.
	A detailed two-photon characterization was reported in Ref. \cite{MPEQDBS18}.

	Spatial mode-mismatch between the mode of the pump laser and the waveguide mode limits the coupling efficiency to roughly $30\%$.
	This bounds the maximum pulse energy inside the waveguide and, hence, the maximum mean photon number, i.e., the squeezing that can be generated.
	This creates spatial mode mismatch between single-mode waveguide and the laser's mode, limiting our pump coupling to $30\%$, which also limits access to high mean photon numbers from the source.
	Broadband filters ($8\,\mathrm{nm}$ and $10\,\mathrm{nm}$) are placed on both arms to avoid the side lobes coming from the quasi-phase-matching cardinal sine ($\mathrm{sinc}$) function.
	We observed a polarization-based Hong--Ou--Mandel dip visibility of $\approx93\%$ between signal and idler photons.
	Reduction in visibility can be caused by the reason following:
		uncertainty in the birefringence compensation crystal length,
		non-perfect wave plates,
		and asymmetry of the photons.

	Since we are working with a TMD, we introduced a pulse-picking device based on a high-voltage electro-optic modulator in combination with a polarizing beam splitter (PBS).
	We reduce the repetition of the laser to $1\,\mathrm{MHz}$ in a synchronized manner with the laser system and the detection scheme.
	A global electronic clock signal is derived from the laser.
	This signal is then split by a clock synthesizer to have multiple phase-locked clock signal to control pulse picking and data collection.
	Gating the picked pulses permits to create
	large enough time bins to accommodate clicks between trigger gates.

	A TMD setup with $75\%$ transmission efficiency is used for photon-number-resolving measurements; see Fig. 1(b) in the Letter.
	At each splitting, a $2\times2$ fiber beam splitter is used.
	To compensate the deadtime of our nanowire detectors, low-loss fibers are spliced with fiber beam splitters, designed to deliver a $100\,\mathrm{ns}$ time-bin seperation.
	Superconducting nanowire single-photon detectors (SNSPD) are used, with $80\%$ detection efficiency and between $50\,\mathrm{ns}$ to $60\,\mathrm{ns}$ deadtime.
	Since we are using long fibers in our experiments, polarization controllers are placed before the TMD input and before the SNSPD detectors to maximize efficiencies.
	Motorized rotation stages are used for different wave-plate settings.

\section{Comparing noise sensitivity of the linear and nonlinear approach}
\label{sec:CompareLandNL}

	For comparing linear and nonlinear polarization squeezing conditions, we can focus on one specific direction on the Poincar\'e sphere, without a loss of generality.
	Also, we set the quantum efficiency $\eta=1$ as we want to explore noise robustness rather than losses.
	With those premises, the linear Stokes operator we shall consider is
	\begin{equation}
		\hat S_\mathrm{L}=\hat a^\dag\hat a-\hat b^\dag\hat b,
	\end{equation}
	pointing along the $z$ direction.
	And the nonlinear squeezing operator is
	\begin{equation}
	\begin{aligned}
		\hat S_\mathrm{NL}
		=&N\left(\hat 1-{:}e^{-\hat a^\dag\hat a/N}{:}\right)
		-N\left(\hat 1-{:}e^{-\hat b^\dag\hat b/N}{:}\right)
		\\
		=&N\left(1-\frac{1}{N}\right)^{\hat b^\dag\hat b}
		-N\left(1-\frac{1}{N}\right)^{\hat a^\dag\hat a},
	\end{aligned}
	\end{equation}
	where we applied the identity ${:}e^{-x\hat n}{:}=(1-x)^{\hat n}$ for a photon-number operator $\hat n$ to rewrite the definition.

	Now, nonclassicality criteria may be formulated via normally ordered variances of the linear and nonlinear Stokes operators.
	Specifically, for classical light, the following bounds are obeyed:
	\begin{equation}
		0\stackrel{\text{cl.}}{\leq}\langle{:}\hat S_\mathrm{L}^2{:}\rangle-\langle{:}\hat S_\mathrm{L}{:}\rangle^2
		\quad\text{and}\quad
		0\stackrel{\text{cl.}}{\leq}\langle{:}\hat S_\mathrm{NL}^2{:}\rangle-\langle{:}\hat S_\mathrm{NL}{:}\rangle^2.
	\end{equation}
	The latter inequality was introduced and experimentally probed in our paper.
	The former constraint can be recast into the more common form
	$
		\langle \hat S_0\rangle\stackrel{\text{cl.}}{\leq}\langle(\Delta\hat S_\mathrm{L})^2\rangle
	$,
	where $\hat S_0=\hat a^\dag\hat a+\hat b^\dag\hat b$,
	by resolving the normal ordering using fundamental commutation relations.

	As an archetypal model of noise, we consider a convolution of both polarizations with thermal background, i.e., a convolution with the Gaussian phase-space distribution $(\pi\bar n)^{-2}e^{-(|\alpha|^2+|\beta|^2)/\bar n}$.
	For a single mode in Glauber-Sudarshan representation, we thus obtain
	\begin{equation}
	\begin{aligned}
		\langle{:}
			f(\hat a^\dag,\hat a)
		{:}\rangle_{\bar n}
		=&\int d^2\alpha \int d^2\beta\,P(\beta)\frac{e^{-|\alpha-\beta|^2/\bar n}}{\pi\bar n} f(\alpha^\ast,\alpha)
		\\
		=&\int d^2\gamma\frac{e^{-|\gamma|^2/\bar n}}{\pi\bar n}
		\langle{:}f(\hat a^\dag+\gamma^\ast,\hat a+\gamma){:}\rangle,
	\end{aligned}
	\end{equation}
	for arbitrary operator functions $f$.
	Therein, the subscript $\bar n$ on the left-hand side indicates the noisy expectation value.
	Using Gaussian integral identities enables us to compute how photon-number-based observables transform under this noise.
	Specifically, we find
	\begin{equation}
	\begin{aligned}
		\langle{:}\hat n^k{:}\rangle_{\bar n}
		=&\sum_{j=0}^k\binom{k}{j}\frac{k!}{(k-j)!}\bar n^j\langle{:}\hat n^{k-j}{:}\rangle
		\\
		\text{and}\quad
		\langle{:}e^{-z\hat n}{:}\rangle_{\bar n}
		=&\frac{1}{1+\bar n z}\langle{:}
			e^{-z\hat n/(1+\bar nz)}
		\rangle{:}\rangle.
	\end{aligned}
	\end{equation}
	As this holds true for all states, we can say equivalently that
	${:}e^{-z\hat n}{:}\mapsto {:}e^{-z\hat n/(1+\bar n z)}{:}/(1+\bar n z)$, $\hat n\mapsto\hat n+\bar n$, and ${:}\hat n^2{:}\mapsto {:}\hat n^2{:}+4\bar n\hat n+2\bar n^2$ hold true, for the relevant values $k=1,2$.

	Using the above input-output relations for noise additions and extending them to two modes, we can express our nonclassicality criteria to include such noise contributions as well.
	In particular, we have
	\begin{equation}
	\begin{aligned}
		\langle{:}\hat S_\mathrm{L}{:}\rangle_{\bar n}
		=&\langle\hat a^\dag\hat a\rangle-\langle \hat b^\dag\hat b\rangle
		\quad\text{and}
		\\
		\langle{:}\hat S_\mathrm{L}^2{:}\rangle_{\bar n}
		=&\langle(\hat a^\dag\hat a)^2\rangle
		-2\langle\hat a^\dag\hat b\hat a^\dag\hat b\rangle
		+\langle(\hat b^\dag\hat b)^2\rangle
		\\
		&+(2\bar n-1)\left(
			\langle\hat a^\dag\hat a\rangle
			+\langle\hat b^\dag\hat b\rangle
		\right)
		+2\bar n^2
	\end{aligned}
	\end{equation}
	for the linear case.
	And, for nonlinear Stokes operators, the noise model yields
	\begin{equation}
	\begin{aligned}
		\langle{:}\hat S_\mathrm{NL}{:}\rangle_{\bar n}
		=&\frac{N}{1+\bar n/ N}
			\left\langle\left[\frac{N+\bar n-1}{N+\bar n}\right]^{\hat b^\dag\hat b}\right\rangle
		\\
		&-\frac{N}{1+\bar n/ N}
			\left\langle\left[\frac{N+\bar n-1}{N+\bar n}\right]^{\hat a^\dag\hat a}\right\rangle
	\end{aligned}
	\end{equation}
	and
	\begin{equation}
	\begin{aligned}
		\langle{:}\hat S_\mathrm{NL}^2{:}\rangle_{\bar n}
		=&\frac{N^2}{1+2\bar n/ N}
			\left\langle\left[\frac{N+2\bar n-2}{N+2\bar n}\right]^{\hat b^\dag\hat b}\right\rangle
		\\
		&-\frac{2N^2}{(1+\bar n/ N)^2}
			\left\langle\left[\frac{N+\bar n-1}{N+\bar n}\right]^{\hat a^\dag\hat a+\hat b^\dag\hat b}\right\rangle
		\\
		&+\frac{N^2}{1+2\bar n/ N}
			\left\langle\left[\frac{N+2\bar n-2}{N+2\bar n}\right]^{\hat a^\dag\hat a}\right\rangle.
	\end{aligned}
	\end{equation}
	With this calculation for describing linear and nonlinear first- and second-order noisy moments, we can now analyze the resilience of the polarisation-squeezing criteria in the presence of noise.

\begin{figure}
	\includegraphics[width=.7\columnwidth]{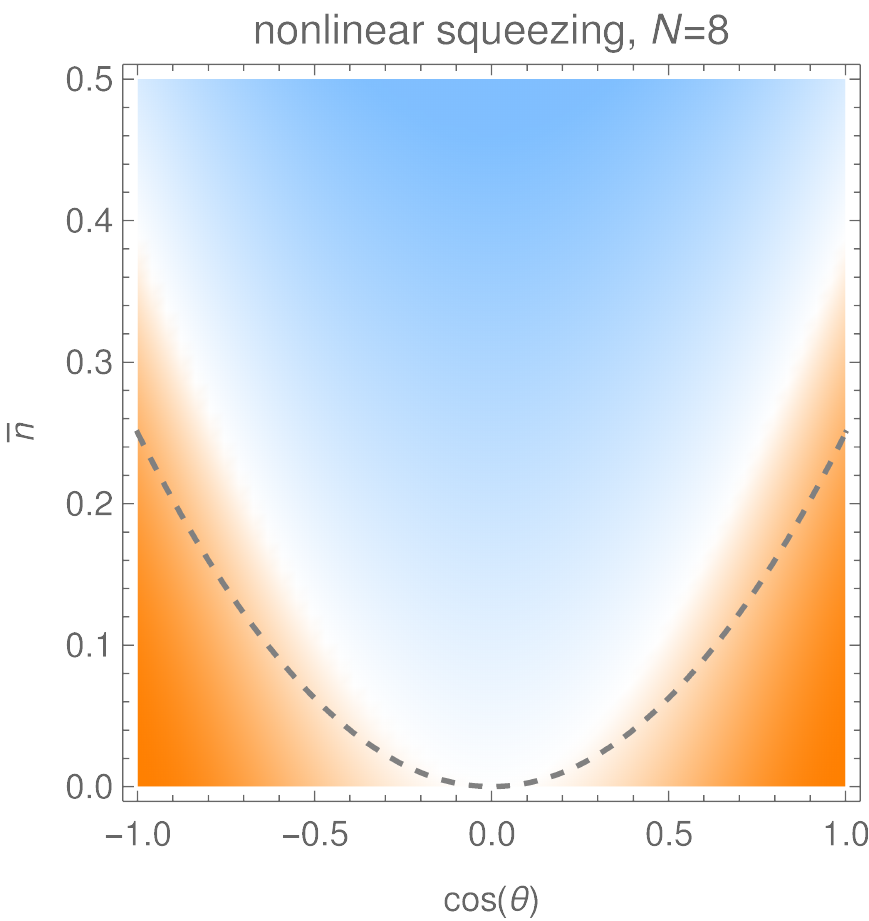}
	\hfill
	\includegraphics[width=.12\columnwidth]{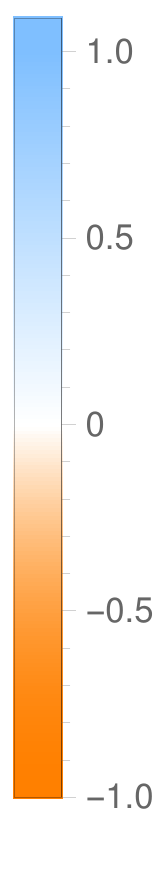}
	\caption{%
		Nonlinear polarization squeezing, $\langle{:}(\Delta\hat S_\mathrm{NL})^2{:}\rangle_{\bar n}<0$, for the state in Eq. \eqref{eq:SMprobestate} with thermal-noise background that is defined through the mean thermal photon number $\bar n$.
		Negative (orange) values indicate detected nonclassicality with our new criterion.
		All values below dashed line also show linear polarization squeezing, but not above this threshold.
		We chose $N=8$ detection bins for the plot as this corresponds to our experiment.
	}\label{fig:SMcompare}
\end{figure}

	For this comparison, we take the simplest example of a single-photon with arbitrary polarization that is subjected to the aforementioned Gaussian noise.
	The ideal state reads
	\begin{equation}
		\label{eq:SMprobestate}
		|\Psi\rangle
		=\cos\tfrac{\theta}{2}|1\rangle\otimes|0\rangle
		+e^{i\phi}\sin\tfrac{\theta}{2}|0\rangle\otimes|1\rangle
	\end{equation}
	in the two-mode polarization photon-number basis.
	Note that the occurring angles can be also used to alter the measurement direction;
	in particular, $\theta$ determines the orientation of the photon's polarization with respect to the $z$ axis of the Poincar\'e sphere.
	For this state, we obtain $\langle{:}\hat S_\mathrm{L}{:}\rangle_{\bar n}=\cos\theta$ and $\langle{:}\hat S^2_\mathrm{L}{:}\rangle_{\bar n}=4\bar n$, resulting in $\langle{:}(\Delta\hat S_\mathrm{L})^2{:}\rangle_{\bar n}=4\bar n-\cos^2\theta$.
	Therefore, the normally ordered variance exhibits negativities, i.e., linear polarization squeezing, for $\bar n<[\cos \theta]^2/4$.
	For the nonlinear scenario, we find $\langle{:}\hat S_\mathrm{NL}{:}\rangle_{\bar n}=\cos\theta/[1+\bar n/N]^2$ and $\langle{:}\hat S^2_\mathrm{NL}{:}\rangle_{\bar n}=\tfrac{2N(N+2\bar n-1)}{[1+2\bar n/N]^2}-\tfrac{2N(N+\bar n-1)}{[1+\bar n/N]^3}$.
	The resulting normally ordered, nonlinear, and noisy variance is shown in Fig. \ref{fig:SMcompare}.

	When compared with the linear case (dashed boundary in Fig. \ref{fig:SMcompare}), nonlinear squeezing persists for higher noise contributions $\bar n$ than possible for its linear counterpart, excluding $\cos\theta=0$.
	Therefore, our criterion is generally superior to---i.e., more robust than---the previously known linear squeezing conditions in this regime.
	For instance for $\cos\theta=\pm1$, we find an upper noise bound of $0.25$ for the linear case while the threshold is at $\approx 0.385$ for the nonlinear scenario, an improvement of more than $54\%$ for the depicted $N=8$ click detection bins.
	(For comparison, $N=2$ and $N=128$ respectively result in a $75\%$ and $47\%$ increased noise robustness.)
	Also, the single-photon states on a thermal background for specific $\bar n$ values can be seen as states for which linear polarization squeezing does not apply, but nonlinear squeezing can be certified by our approach.

\section{Theoretical model for the experiment}
\label{secSM:Model}

	In this section, we provide the fully exact calculation that models our experiment (as depicted in the main paper) and that is used to compare our data with.
	This is done in multiple steps.
	First, a phase-space model is formulated.
	Second, the source is considered.
	Third, the click-counting outcomes are determined.
	Finally, all components are put together.

\subsection{Phase-space description of two-mode squeezed vacuum}

	The PDC produces two-mode squeezed vacuum states,
	\begin{equation}
		|\lambda\rangle=\sqrt{1-|\lambda|^2}\sum_{n=0}^\infty \lambda^n |n,n\rangle,
	\end{equation}
	with $|\lambda|<1$.
	For deriving the $P$ function, we express $|n\rangle$ in terms of coherent states in a convenient manner.
	To this end, the relation
	\begin{equation}
		e^{-|\alpha|^2/2}\partial_\alpha^n\left(e^{|\alpha|^2/2}|\alpha\rangle\right)=\hat a^{\dag n}|\alpha\rangle
	\end{equation}
	is useful.
	Namely, it provides an expression for $|m\rangle\langle n|$ to decompose the density operator;
	this expression reads
	$
		|m\rangle\langle n|=\frac{1}{\sqrt{m!n!}}\left.
			e^{-|\alpha|^2}\partial_\alpha^m\partial_{\alpha^\ast}^n\left(
				e^{|\alpha|^2}|\alpha\rangle\langle\alpha|
			\right)
		\right|_{\alpha=0}
	$.
	Thus, we have
	\begin{equation}
	\begin{aligned}
		&|\lambda\rangle\langle\lambda|
		\\
		=&\sum_{m,n=0}^\infty
		\int d^2\alpha\,d^2\beta\,
		(1-|\lambda|^2)\lambda^{m}\lambda^{\ast n}\delta(\alpha)\delta(\beta)
		\\
		&\times 
		\frac{e^{-|\alpha|^2-|\beta|^2}}{m!n!}
		\partial_\alpha^m\partial_{\alpha^\ast}^n\partial_\beta^m\partial_{\beta^\ast}^n\left(
			e^{|\alpha|^2+|\beta|^2}|\alpha,\beta\rangle\langle\alpha,\beta|
		\right)
		\\
		=&\int d^2\alpha\,d^2\beta\,P_{\lambda}(\alpha,\beta)|\alpha,\beta\rangle\langle\alpha,\beta|,
	\end{aligned}
	\end{equation}
	with
	$
		P_{\lambda}(\alpha,\beta)=
		e^{|\alpha|^2+|\beta|^2} \exp[{\lambda \partial_\alpha \partial_\beta + \lambda^\ast \partial_{\alpha^\ast} \partial_{\beta^\ast}}]
		\times
		[e^{-|\alpha|^2-|\beta|^2}\delta(\alpha)\delta(\beta)]
	$
	that is obtained from integration by parts.
	Thus, we can proceed with a semiclassical picture in which $|\alpha,\beta\rangle$ are propagated.

\subsection{Generation of macroscopic Bell states}

	For forward (clockwise) and backward (counterclockwise) pumping of the PDC source, we get two two-mode squeezed vacuum states.
	In our semiclassical picture, this translates to coherent states $|\alpha,\beta,\gamma,\delta\rangle$---where $\alpha$ is the forward signal in horizontal polarization, $\beta$ is the forward idler in vertical polarization, $\gamma$ is the backward signal in horizontal polarization, and $\delta$ is the backward idler.
	Jointly, we describe this via the four-mode coherent state
	$
		|\alpha,\beta,\gamma,\delta\rangle
	$.

	As depicted in the setup overview in the paper, in the first step, the counterclockwise light is polarization swapped, resulting in $|\alpha,\beta,\delta,\gamma\rangle$.
	The Sagnac PBS then shuffles the polarization of forward and backward light from the loop, $|\alpha,\gamma,\delta,\beta\rangle$, where the first and last two components respectively belong to arm $A$ and $B$.
	The phase controller in arm $A$ produces $|\alpha,\gamma e^{i\phi},\delta,\beta\rangle$.
	Hence, the generated state reads
	\begin{equation}
	\begin{aligned}
		\hat\rho
		=&\int d^2\alpha\,d^2\beta\,d^2\gamma\,d^2\delta\,
		P_\lambda(\alpha,\beta)
		\\
		&\times P_\lambda(\gamma,\delta)
		|\alpha,\gamma e^{i\phi},\delta,\beta\rangle\langle \alpha,\gamma e^{i\phi},\delta,\beta|
		\\
		=&\int d^2\alpha'\,d^2\beta'\,d^2\gamma'\,d^2\delta'\,
		P_\lambda(\alpha',\delta')
		\\
		&\times P_\lambda(\beta'e^{-i\phi},\gamma')
		|\alpha',\beta',\gamma',\delta'\rangle
		\langle \alpha',\beta',\gamma',\delta'|
		\\
		=&(1-|\lambda|^2)^2\sum_{m_1,n_1,m_2,n_2=0}^\infty
		\lambda^{m_1}\lambda^{\ast n_1}\left(\lambda e^{i\phi}\right)^{m_2}
		\\
		&\times
		\left(\lambda e^{i\phi}\right)^{\ast n_2}
		|m_1,m_2,m_2,m_1\rangle
		\langle n_1,n_2,n_2,n_1|.
	\end{aligned}
	\end{equation}
	Since this is a pure state, we can also use the form
	$
		|\psi\rangle=(1-|\lambda|^2)\sum_{m,n=0}^\infty \lambda^m
		\left(\lambda e^{i\phi}\right)^n|m,n,n,m\rangle
	$,
	as given in the paper.
	If convenient,  we can additionally write the state as $|\psi\rangle=(1-|\lambda|^2)\exp(\lambda \hat a^\dag\hat d^\dag+\lambda e^{i\phi} \hat b^\dag\hat c^\dag)|0,0,0,0\rangle$.

\subsection{Click-counting detection}

	Regardless of the studied quantity, be it joint probabilities or moments, the TMD measurement is given in terms of normally-ordered exponentials,
	\begin{equation}
		\hat M={:}\exp\left(
			-X_a \hat a^\dag\hat a
			-X_b \hat b^\dag\hat b
			-X_c \hat c^\dag\hat c
			-X_d \hat d^\dag\hat d
		\right){:}
	\end{equation}
	In this context, it is worth mentioning the useful identity
	$
		{:}e^{-\hat a^\dag\hat a/[\bar n+1]}{:}/[\bar n+1]
		=(\pi\bar n)^{-1}\int d^2\alpha\,e^{-|\alpha|^2/\bar n}|\alpha\rangle\langle \alpha|,
	$
	which holds true for all $\bar n>0$.
	Thus, we can write
	\begin{equation}
	\begin{aligned}
		\hat M=&\int \frac{
			d^2\alpha\,d^2\beta\,d^2\gamma\,d^2\delta\,
		}{
			\pi^4(1-X_a)(1-X_b)(1-X_c)(1-X_d)
		}
		\\
		&\times\exp\left(
			-\frac{X_a|\alpha|^2}{1-X_a}
			-\frac{X_b|\beta|^2}{1-X_b}
			-\frac{X_c|\gamma|^2}{1-X_c}
			-\frac{X_d|\delta|^2}{1-X_d}
		\right)
		\\
		&\times
		|\alpha,\beta,\gamma,\delta\rangle\langle \alpha,\beta,\gamma,\delta|.
	\end{aligned}
	\end{equation}

	Since we only measure the first mode in arm $A$ and the second mode in $B$, we can fix $X_b=0$ and $X_c=0$.
	Furthermore, the combination of waveplates, being set to identical values in both arms, superimposes the fields.
	With that information, the measurement operator takes the modified form
	\begin{equation}
	\begin{aligned}
		\hat M=&\int \frac{
			d^2\alpha\,d^2\beta\,d^2\gamma\,d^2\delta\,
		}{
			\pi^4(1-X_a)(1-X_d)
		}
		\\
		&\times\exp\left(
			-\frac{X_a|\tau\alpha+\rho\beta|^2}{1-X_a}
			-\frac{X_d|-\rho^\ast\gamma+\tau^\ast\delta|^2}{1-X_d}
		\right)
		\\
		&\times
		|\alpha,\beta,\gamma,\delta\rangle\langle \alpha,\beta,\gamma,\delta|.
	\end{aligned}
	\end{equation}

	Finally, we compute the sought-after expectation value.
	For this purpose, the following integral formula is useful,
	\begin{equation}
	\begin{aligned}
		&\int_{\mathbb C^n} d^{2n}\vec\alpha\,\exp\left(
			-\vec \alpha^\dag Y\vec\alpha
			+\vec \alpha^\mathrm{T} Z\vec\alpha
			+\vec \alpha^\dag Z^\ast\vec\alpha^\ast
		\right)
		\\
		=&\pi^n\left(\det\begin{bmatrix}
			Y & 2Z^\ast \\ 2Z & Y^\ast
		\end{bmatrix}\right)^{-1/2},
	\end{aligned}
	\end{equation}
	with $Y=Y^\dag$, $Z=Z^\mathrm{T}$, and $\left[\begin{smallmatrix}Y&2Z^\ast\\2Z&Y^\ast\end{smallmatrix}\right]>0$.
	Applying that and using $\vec \alpha=\left[\begin{smallmatrix} \alpha \\ \beta \\ \gamma \\ \delta \end{smallmatrix}\right]$, we find
	\begin{equation}
	\begin{aligned}
		&\langle\psi|\hat M|\psi\rangle
		\\
		=&\frac{(1+|\lambda|^2)^2}{\pi^4(1-X_a)(1-X_d)}
		\\&\times\int d^{2\cdot4}\vec\alpha\,
		\exp\left(
			-\vec \alpha^\dag Y\vec\alpha
			+\vec \alpha^\mathrm{T} Z\vec\alpha
			+\vec \alpha^\dag Z^\ast\vec\alpha^\ast
		\right)
		\\
		=&\frac{(1-|\lambda|^2)^2(1-X_a)(1-X_d)}{\det\left[\begin{smallmatrix}
			1-|\rho|^2 X_a & \rho\tau^\ast X_a & 0 & \tilde\lambda \\
			\rho^\ast\tau X_a & 1-|\tau|^2 X_a & \tilde\lambda e^{i\phi} & 0 \\
			0 & \tilde\lambda^\ast e^{-i\phi} & 1-|\tau|^2 X_d & -\rho^\ast\tau X_d \\
			\tilde\lambda^\ast & 0 & -\rho\tau^\ast X_d & 1-|\rho|^2 X_d
		\end{smallmatrix}\right]},
	\end{aligned}
	\end{equation}
	with the modified parameter $\tilde \lambda=\sqrt{(1-X_a)(1-X_d)}\lambda$ as well as $
		2Z^\ast=\lambda\left[\begin{smallmatrix}
			0 & 0 & 0 & 1
			\\
			0 & 0 & \exp(i\phi) & 0
			\\
			0 & \exp(i\phi) & 0 & 0
			\\
			1 & 0 & 0 & 0
		\end{smallmatrix}\right]
	$ and $
		Y=\left[\begin{smallmatrix}
			Y_{1,1} & Y_{1,2}
			\\
			Y_{2,1} & Y_{2,2}
		\end{smallmatrix}\right]
	$, where
	$
		Y_{1,1}=\left[\begin{smallmatrix}
			(1-|\rho|^2X_a)/(1-X_a) & (\tau^\ast\rho X_a)/(1-X_a)
			\\
			(\rho^\ast\tau X_a)/(1-X_a) & (1-|\tau|^2X_a)/(1-X_a)
		\end{smallmatrix}\right]
	$,
	$
		Y_{2,2}=\left[\begin{smallmatrix}
			(1-|\tau|^2X_d)/(1-X_d) & (-\tau^\ast\rho X_d)/(1-X_d)
			\\
			(-\rho^\ast\tau X_d)/(1-X_d) & (1-|\rho|^2X_d)/(1-X_d)
		\end{smallmatrix}\right]
	$, and
	$
		Y_{1,2}=Y_{2,1}=\left[\begin{smallmatrix}
			0 & 0
			\\
			0 & 0
		\end{smallmatrix}\right]
	$.

\subsection{Application to experiment at hand}

	We conclude by using the above results for our specific scenario.
	Firstly, we have $e^{i\phi}=\pm 1$.
	Secondly, QWP and HWP are respectively defined through the Jones matrices
	\begin{equation}
		\begin{bmatrix}
			\tau_\mathrm{\tiny QWP} & \rho_\mathrm{\tiny QWP}
			\\
			-\rho_\mathrm{\tiny QWP}^\ast & \tau_\mathrm{\tiny QWP}^\ast
		\end{bmatrix}
		=\begin{bmatrix}
			\frac{1-i\cos(2\theta)}{\sqrt2} & \frac{-i\sin(2\theta)}{\sqrt2}
			\\
			\frac{-i\sin(2\theta)}{\sqrt2} & \frac{1+i\cos(2\theta)}{\sqrt2}
		\end{bmatrix}
	\end{equation}
	and
	\begin{equation}
		\begin{bmatrix}
			\tau_\mathrm{\tiny HWP} & \rho_\mathrm{\tiny HWP}
			\\
			-\rho_\mathrm{\tiny HWP}^\ast & \tau_\mathrm{\tiny HWP}^\ast
		\end{bmatrix}
		=\begin{bmatrix}
			-i\cos(2\theta) & -i\sin(2\theta)
			\\
			-i\sin(2\theta) & +i\cos(2\theta)
		\end{bmatrix}.
	\end{equation}
	In addition, the moments and statistics from the click-counting device allow for values $X_a\in \left\{\frac{\eta m_A}{N}: m_A=0,\ldots,N\right\}$ and $X_d\in \left\{\frac{\eta m_B}{N}: m_B=0,\ldots,N\right\}$, where $N$ is the number of detection bins.

	In the first step, this simplifies the previously obtained expectation value to
	\begin{equation}
	\begin{aligned}
		&\langle\psi|\hat M|\psi\rangle
		=\left\langle{:}
			\exp\left(
				-\frac{\eta m_A}{N}\hat n_A
				-\frac{\eta m_B}{N}\hat n_B
			\right)
		{:}\right\rangle
		\\
		=&\frac{
			N^2
			(1-|\lambda|^2)^2
			(N-\eta m_A)
			(N-\eta m_B)
		}{\det\left[\begin{smallmatrix}
			N-|\rho|^2 \eta m_A & \rho\tau^\ast \eta m_A & 0 & \lambda^\prime \\
			\rho^\ast\tau \eta m_A & N-|\tau|^2 \eta m_A & \pm \lambda^\prime & 0 \\
			0 & \pm \lambda^{\prime\ast} & N-|\tau|^2 \eta m_B & -\rho^\ast\tau \eta m_B \\
			\lambda^{\prime\ast} & 0 & -\rho\tau^\ast \eta m_B & N-|\rho|^2 \eta m_B
		\end{smallmatrix}\right]},
		\\
		\\
	\end{aligned}
	\end{equation}
	where $\lambda^\prime=\sqrt{(N-\eta m_A)(N-\eta m_B)}\lambda$ and photon-number operators $\hat n_A=\hat a^\dag\hat a$ and $\hat n_B=\hat b^\dag\hat b$.

	We can now consider the nonclassicality criteria, beginning with the second-order one.
	The normally ordered variance expands as
	\begin{equation}
	\begin{aligned}
		&\langle{:}
			(\Delta \hat S_\mathrm{NL})^2
		{:}\rangle
		\\
		=&N^2\left(
			\langle{:}e^{-2\eta\hat n_A/N}{:}\rangle
			-\langle{:}e^{-\eta\hat n_A/N}{:}\rangle^2
		\right.
		\\
		&\left.
			+\langle{:}e^{-2\eta\hat n_B/N}{:}\rangle
			-\langle{:}e^{-\eta\hat n_B/N}{:}\rangle^2
		\right.
		\\
		&\left.
			-2\langle{:}e^{-\eta(\hat n_A+\hat n_B)/N}{:}\rangle
			+2\langle{:}e^{-\eta\hat n_A/N}{:}\rangle\langle{:}e^{-\eta\hat n_B/N}{:}\rangle
		\right)
	\end{aligned}
	\end{equation}
	using $\hat S_\mathrm{NL}=N{:}\exp(-\tfrac{\eta}{N}\hat n_B){:}-N{:}\exp(-\tfrac{\eta}{N}\hat n_A){:}$.
	Secondly, the matrix of normally-ordered moments is defined in terms of elements
	\begin{equation}
	\begin{aligned}
		&\langle{:}\pi_A^{m_A+n_A}\pi_B^{m_B+n_B}{:}\rangle
		\\
		=&\sum_{k_A=0}^{m_A+n_A}\sum_{k_B=0}^{m_B+n_B}
		\binom{m_A+n_A}{k_A}\binom{m_B+n_B}{k_B}
		\\
		&\times (-1)^{k_A+k_B}
		\langle{:}
			e^{-\eta(k_A\hat n_A+k_B\hat n_B)/N}
		{:}\rangle,
	\end{aligned}
	\end{equation}
	applying $\hat\pi_{S}=\hat 1-{:}\exp(-\eta\hat n_S/N){:}$ for $S\in\{A,B\}$.
	To those we apply the same analytical formulas from above.